\documentclass[12pt]{article}
\newcommand{\be}{\begin{equation}}
\newcommand{\ee}{\end{equation}}
\newcommand{\bea}{\begin{eqnarray}}
\newcommand{\eea}{\end{eqnarray}}
\newcommand{\nn}{\nonumber}
\newcommand{\lb}{\left[}
\newcommand{\rb}{\right]}
\newcommand{\ac}{\mathcal{A}}
\newcommand{\bc}{\mathcal{B}}
\newcommand{\cc}{\mathcal{C}}
\newcommand{\hf}{{1\over 2}}
\newcommand{\qq}{{\underline q}}
\newcommand{\pp}{{\underline p}}
\newcommand{\pki}{\stackrel{k-1}{\Pi}{}\!\!}
\newcommand{\dd}{\stackrel{\leftarrow}{\partial_t}{}\!\!}

\begin{document}
\begin{titlepage}

\title{On the structure of the constraint algebra for 
systems whose gauge transformations depend on
higher order time derivatives of the gauge parameters}

\author{M.N. Stoilov\\
{\small\it Bulgarian Academy of Sciences,}\\
{\small\it Institute of Nuclear Research and Nuclear Energy,}\\
{\small\it Blvd. Tzarigradsko Chausse\'e 72, Sofia 1784, Bulgaria}\\
{\small e-mail: mstoilov@inrne.bas.bg}}

\maketitle     

\begin{abstract}
The dynamical systems invariant under gauge transformations
with higher order time derivatives of the gauge parameter
are considered from the Hamiltonian point of view.
We investigate the consequences of the basic requirements that
the constraints on the one hand 
and the  Hamiltonian and constraints on the other hand
form two closed algebras.
It is demonstrated that these simple algebraic requirements
lead to rigid relations in the constraint algebra. 
\end{abstract}
              

key words: higher stage constraints

\end{titlepage}

\section*{Introduction}

Dynamical systems in which the gauge transformations involve
higher order derivatives of the gauge parameters are considered in the literature
both from the Lagrangean and Hamiltonian point of view \cite{N}--\cite{De}.
In the Lagrangean approach the corresponding
Noether identities are obtained
and in the Hamiltonian approach the constraints generating the
gauge transformations are constructed.
In the present paper we address
some aspects of the possible representations of the
constraint algebra in the Hamiltonian approach.

Any dynamical system 
with gauge symmetry  is characterized in the Hamiltonian approach by its
 Hamiltonian $H$ and constraints $\varphi_a, \;\;a=1,\dots,c$.
 The Hamiltonian and constraints
are functions of the phase space variables
$q_m$ and $p_m,\;\;m=1,\dots,n$.
The constraints generate the gauge transformation 
of any phase space dynamical quantity $g(p,q)$
through the Poisson bracket relations
\be
\delta_\epsilon g = \epsilon_a \lb g, \varphi_a\rb. \label{gt}
\ee
The Hamiltonian generate  (again through the Poisson bracket relations)
the time evolution (up to a gauge transformation) of any $g$
\be
 \dot g \equiv   {d g\over d t} = \lb g,H\rb. \label{dyn}
\ee
In eq.(\ref{gt})
the parameters of the gauge transformation $\epsilon_a$ can be 
arbitrary functions of the time $t$.
Due to the specific character of the time in the Hamiltonian approach
there are no time derivatives of any order of $\epsilon_a$
in the transformation (\ref{gt}).
On the other hand, if we consider a field theory it is possible to 
have spatial derivatives acting on $\epsilon_a$.

It is not possible to pick up arbitrary Hamiltonian and
constraints and to obtain a well defined dynamical model.
There are some consistency conditions which the Hamiltonian and constraints
have to satisfy. 
First, the commutator of two gauge transformations has to be 
a gauge transformation.
Together with  the Jacobi identity this means that constraints 
form a closed gauge algebra with respect to the Poisson bracket relations 
\be
\lb\varphi_a, \varphi_b\rb = C_{abe}\varphi_e. \label{algi}
\ee
Here $C_{abc}$ are the structure functions of the gauge algebra.
Second, the time evolution has to preserve the gauge algebra (\ref{algi}).
In other words the Hamiltonian and constraints also have to form
a closed algebra, i.e. besides eq.(\ref{algi}) the following relation
has to be satisfied as well
\be
\lb H, \varphi_a\rb  =  U_{ab}\varphi_b.\label{algii}
\ee
In the simplest but very common case the structure functions
$C_{abe}$ and  $U_{ab}$ do not depend on the dynamical variables.
In this case both the gauge algebra (\ref{algi})
and the algebra of the constraints and Hamiltonian
(\ref{algi},\ref{algii}) are Lie algebras.
If the gauge algebra is semi-simple or Abelian
then the structure constants $C_{abe}$ only matters for the algebra
of the Hamiltonian and constraints.
The structure constants $U_{ab}$ are not important because they
are due to weakly zero terms
(terms proportional to the constraints) in the Hamiltonian \cite{M}.
Such terms can be freely removed from the Hamiltonian
(thus obtaining the so called `canonical Hamiltonian')
and if we do so, we get that $U_{ab}$ are zeros.
In other words, the canonical Hamiltonian is always gauge invariant.

The Hamiltonian approach to the constraint systems is
equivalent to the first order Lagrangean approach
with the following Lagrangean.
\be
L=p\dot q - H - \lambda_a \varphi_a .\label{lag}
\ee
Here $\lambda_a$ are the Lagrange multipliers.
Their gauge transformation is given below:
\be
\delta_\epsilon \lambda_a = \partial_t\epsilon_a  +
\epsilon_c C_{cba}\lambda_b. \label{varlm}
\ee
Note that we have a term with time derivative of the gauge parameters 
in eq.(\ref{varlm}).
It has been already stressed that the Hamiltonian approach does not
allow time derivatives of the gauge parameter.
Therefore, we need some modification of this approach
if we want to handle within it the gauge transformation of the 
Lagrange multipliers. 
It is shown in Ref.\cite{M} that the transformation (\ref{varlm})
can be generated by the following constraints  which act in the
phase space of Lagrange multipliers $\lambda_a$ 
and their momenta $\pi_a$
\be
\hat\varphi_a= \dd\pi_a  + 
\lambda_b C_{abc}\pi_c. \label{lmc}
\ee
Here we introduce the operator of the time derivative $\dd$ which
acts on the gauge parameters and not on the phase space variables. 
If we do not use the canonical Hamiltonian  then in eqs.(\ref{varlm},\ref{lmc})
some extra terms proportional to $U_{ab}$ appear.

Eq.(\ref{varlm}) is an example of a gauge transformation
with first order time derivative of the gauge parameter. 
This example gives us grounds to ask the question  
is it possible, e.g.  in the second order Lagrangean formalism, to
have dynamical variables whose gauge transformation involves
higher than first order time derivatives of the gauge parameter?
The answer of this question is positive.
The aim of the present paper is to investigate
the algebra of constraints which generates gauge transformations 
with higher time derivatives of the gauge parameter.
Some aspects of this problem are considered in \cite{N},\cite{DE}.
Here we focus our attention on the consequences of the required Lie
algebraic structure.
In our investigation we use eq.(\ref{lmc}) as a pattern:
the constraints $\hat\varphi_a$ are polynomials with respect to the 
time derivative operator $\dd$ 
with coefficients  functions in a specific phase space.
We expect the same structure for the generators of the gauge transformations
involving higher order time derivatives of the gauge parameters.
Loosely speaking we shall refer to such gauge transformations 
 as `higher stage' ones.

\section*{Higher stage gauge transformations}

\subsection*{An example}

There is a simple example with higher stage transformations of any finite order.
Consider a mechanical model with $n$ coordinates $q_m, \;\;m=1,\dots,n$ 
with the following Lagrangean
\be
L= \hf\sum_{m=2}^{n} (\dot q_{m-1}-q_m)^2. \label{exL}
\ee
The model is invariant with respect to the following
one parametric gauge transformation:
\be
\delta_\epsilon q_m = \partial_t^{m-1}\epsilon. \label{exT}
\ee
We recall that the parameter $\epsilon$ can be  arbitrary function of the time. 
The Dirac analysis of the Lagrangean (\ref{exL})
shows that we have a primary constraint ($p_n=0$),
a secondary constraint ($p_{n-1}=0$) and so on up to
 $n$-th stage constraint  ($p_1=0$).
All of these constraints are first class.
On the base of this analysis we expect an $n$-parametric gauge symmetry, 
but the symmetry (\ref{exT}) is only one parametric.
Therefore,  none of the primary, secondary and so on constraints
do not generate independent gauge symmetry.
These constraints are projection of the unique gauge symmetry
generator in different subspaces of the phase space 
--- $\{q_n, p_n\},\;\{q_{n-1}, p_{n-1}\}$ and so on.
An interesting feature of the considered model is that the gauge parameter
in the different subspaces is not the same.
As it is seen from eq.(\ref{exT}) the parameter of the gauge transformation
 in the subspace $\{q_m, p_m\}$ is $\partial_t^{m-1}\epsilon$.
An analogy with eqs.(\ref{varlm},\ref{lmc}) suggests that we have to use
the operator of the time derivative acting on the gauge parameter 
when we write down the constraint generating the transformation (\ref{exT}).
The generator which we are looking for is:
\be
\psi = \sum_{m=1}^n \dd^{m-1}p_m
\ee
where $\dd^i$ is the i-th time derivative acting on 
the gauge parameter $\epsilon$.
Note that the coefficients in this series are the different stage constraints
which we obtain through the Dirac prescription.

\subsection*{The general construction}

Hereafter we shall consider only models with finite highest stage
gauge transformations. 
Without this condition the model will be non-local in time.
The general form  of the finite higher stage gauge variation is
\be
\delta_\epsilon q_m = 
\epsilon_a\sum_{i={k_0}}^{k} {\dd^i\over i !}f^i_{a m}. \label{hsgt}
\ee
Here $k_0$ and $k$ are the minimal and maximal order of the 
gauge parameter time derivatives
and $f^i_{am}$ are some (yet unspecified) functions.
Without any loss of generality we can accept that $k_0$ is zero
because the case in which $k_0 \ne 0$ can be brought to the
case $k_0 = 0$ with a redefinition of the parameters $\epsilon_a$, 
such that $\epsilon^{new} = \partial_t^{k_0} \epsilon$.

If the Lagrangean of the model is not with
higher derivatives  then $f_{a m}^i$ are functions of $q$ and $\dot q$
and the symmetry (\ref{hsgt}) can be realized in the phase space
of the model \cite{DE}.
Here we adopt a slightly different approach.
Having in mind eq.(\ref{lmc}),
 we are looking for a realization of the higher stage gauge transformation 
(\ref{hsgt}) in some larger phase space with coordinates $\{\qq,\pp\}$.
This phase space contains besides the initial phase space of the model
also the phase space of the Lagrangean multipliers,
additional phase space variables connected with (possible) higher derivatives
and second class constraints, ghosts, etc.

In general, the enlargement of the phase space requires redefinition of
the constraints and the Hamiltonian.
Terms which live in the new dimensions have to be  added both to the 
initial Hamiltonian and constraints so that the gauge algebra
and the algebra of the Hamiltonian and constraints to remain the same. 
However, if we are using the canonical Hamiltonian there is no need to modify it.
In other words, the canonical Hamiltonian in the enlarged phase space is
a function of the initial phase space variables only.
The reason is the gauge invariance of the canonical Hamiltonian.
On the other hand, if we for some reasons do not use the canonical Hamiltonian
a procedure like the construction of the BRST invariant Hamiltonian
has to be carried out.
Here we assume that we are working with the canonical Hamiltonian.
Therefore, the only things we have to find in the enlarged phase space
are the constraints.
We  are looking for the generators $\psi$ of the transformation (\ref{hsgt})
in the following form:
\be
\psi_a=\sum_{i=0}^{k} {\dd^i\over i !}\varphi^i_a \label{dec}
\ee
where the different stage constraints $\varphi^i_a$ are such
functions in the enlarged phase space so that for any $g(\qq,\pp)$
\be
\delta_\epsilon g = \epsilon_a \lb g, \psi_a\rb. \label{gghs}
\ee
(In the above equation $\lb \;, \rb$ denotes the Poisson brackets in
the $\{\qq,\pp\}$ phase space.)

\subsection*{Consistency conditions}

The basic requirements that the gauge generators on the one hand
and the Hamiltonian and gauge generators on the other hand must
form closed algebras are  valid for any gauge model including the
models with higher stage gauge transformations.
Therefore, for the  higher stage gauge generators the following relation must hold
\be
\lb \psi_a, \psi_b \rb = C_{abc} \psi_c. \label{gcr}
\ee
Hereafter we suppose that the algebra (\ref{gcr}) is a Lie algebra
which we shall denote $\ac$.
The requirement that the time evolution of the constraints
does not produce new constraints leads to the gauge
invariance of the canonical Hamiltonian \cite{M}
in the case of $0$-stage gauge transformations which form Abelian or semi-simple 
Lie algebra. 
The result however
does not depend on the particular realization of the gauge algebra.
So, even for the higher stage gauge transformations
the canonical Hamiltonian has to be gauge invariant, i.e.
\be
\epsilon_a\lb H, \psi_a\rb  =  0.
\ee
From the above equation we get using the arbitrariness of 
the gauge parameters $\epsilon_a$ that
\be
\lb H, \varphi_a^i\rb  =  0\;\;\forall a,i.
\ee

From eqs.(\ref{dec}, \ref{gcr}) and after
 a series expansion  on different powers of $\dd$
we get the Poisson bracket relations between the different 
stage generators $\varphi_a^i$:
\be
\lb \varphi_a^i, \varphi_b^j \rb =
\theta^{k}_{i+j} C_{abc} \varphi_c^{i+j}. \label{cr}
\ee
Here $\theta^i_j$ is the step symbol
\be 
\theta^i_j=\left\{ { \begin{array}{l} 1 \;\;{\rm if }\;\; i \ge j\\
0 \;\;{\rm if } \;\; j>i\end{array}}\right\}
\ee 

Using the fact that $C_{abc}$ are structure constants of a Lie algebra
it is easy to check that the set $\left\{\varphi_a^i\right\}$ generates a Lie 
algebra as well.
This algebra we denote $\bc^{k}$.
It follows from eq.(\ref{cr}) that $\ac$ is a sub-algebra of $\bc^k$.
The algebra $\bc^k$ has the following Killing form:
\be
 g^\bc_{ai\;bj} = \delta_{i0}\delta_{j0} g^\ac_{ab}\label{kf}
\ee
where $g^\ac_{ab}$ is the Killing form of the algebra $\ac$.
Eq.(\ref{kf}) leads to the following Levi--Malcev decomposition of $\bc^k$
in the case when $\ac$ is semi-simple
\be
\bc^k = \cc\;  +)\; \ac. \label{as}
\ee
In the above semi-direct sum decomposition  the algebra $\cc$ is generated by 
$\varphi^i_a$ with $i>0$
while $\ac$ is generated by $\varphi^0_a$.
It turns out that the algebra $\cc$  is not only solvable  but 
it is  nilpotent.

\subsection*{Some representations of the algebra $\bc^k$}

\subsubsection*{Matter representation}

Suppose we know a  representation $\pi(\ac)$ of the algebra $\ac$
acting in a $d$-dimensional vector space $V$.
Then it is possible
 to construct a representation $\Pi(\bc^k)$ of $\bc^k$
in the $(k +1) . d$-dimensional space
 $\stackrel{k +1}{\oplus}V = 
 \underbrace{V\oplus V\oplus\cdots\oplus V}_{k+1}$.
Let us denote by $A_a$ the $d\times d$ matrix representing $\psi_a$ 
(or $\varphi_a^0$)
\be
A_a=\pi(\psi_a) \label{base}
\ee
The  representation  $\Pi(\bc^k)$ is given by  block matrices
such that
\be
\Pi(\varphi_a^m)_{ij}=\theta^k_j\delta^{k+i}_j A_a. \label{sky}
\ee
In eq.(\ref{sky}) the subscripts $i,j = 0,\dots,k$ indicate the block  
row and column position 
 and the block contents is always the matrix $A_a$.
In other worlds, the $\Pi$ representation of $\varphi^0_a$ is given by
a block diagonal matrix  with the matrix $A_a$ in every diagonal block,
and all other blocks equal to zero;
$\Pi(\varphi^1_a)$ is given by a matrix
for which in any block along the block diagonal above the main 
block diagonal sits the matrix $A_a$, all other blocks zero, and so on till
$\Pi(\varphi^k_a)$ for which the only non-zero block is in the
upper right corner where again the matrix $A_a$ sits.

The above construction can be realized in a phase space with coordinates
$\left\{q^0_u,\dots,q^k_v, p^0_u,\dots,p^k_v\right\}$ where 
$u, v = 1,\dots,d$ as follows:
\be
\Pi^m_a = - \sum_{i=0}^{k-m} q^i  A_a  p^{i+m}
\ee
In this realization $q^0$ transforms as a vector (matter),
$q^1$ transforms as $\dot q^0$, 
while the gauge transformations of the other coordinates are more complicated:
\be
\delta_\epsilon q^i = - \sum_{j=0}^i
{1\over j!}\epsilon^{(j)}_a q^{i-j}A_a.
\ee

\subsubsection*{Connection representation}

If the representation $\pi(\ac)$ is the adjoin one 
(and so, $V=\ac$)
then it is possible to construct a representation of the algebra $\bc^k$
in a smaller space, namely in $ \stackrel{k}{\oplus}V$.
Let the matrices $\pki^0_a, \dots,\pki^{k-1}_b$  realize
a representation of the algebra $\bc^{k-1}$ in  
$ \stackrel{k}{\oplus}V$ as described in eq.(\ref{sky}).
Let $T^i_a,\;\;i=0,\dots,k-1$ are the translation generators in 
$ \stackrel{k}{\oplus}V$.
The meaning of the indices $i$ and $a$ of $T^i_a$ is as follows:
the index $i$ indicates the space in the direct sum
and the index $a$ indicates the coordinate 
in this space on which the generator  $T^i_a$ acts.
Note that the dimension of the adjoin representation is $c$
(the number of constraints)
so the range of the translation indices $a$ is correct.
The translation generators satisfy the following commutation relations
\bea
\lb T^i_a, T^j_b \rb &=&0 \;\;\forall i, j\;\&\;\forall a, b\nn\\
\lb \pki^i_a, T^j_b \rb &=& \theta^{k-1}_{i+j} C_{abe} T^{i+j}_e.
\eea
We are looking for linear combinations $\bar\Pi^i_a$ 
of the operators $T^i_a$ and $\pki^i_a$
\bea 
\bar\Pi^0_a   &=& \pki^0_a\nn\\
\bar\Pi^i_a   &=& \pki^i_a + \alpha_i T^{i-1}_a,\;\;i=0,\dots,k-1\nn\\
\bar\Pi^k_a   &=& \alpha_k T^{k-1}_a\label{lr}
\eea
such that $\bar\Pi^i_a,\;\;i=0,\dots,k$ to satisfy the commutator relations of
the algebra $\bc^k$.
The result is that the coefficients $\alpha_i$ have to be such that
\be
\alpha_{i+j}  = \alpha_i + \alpha_j\;\;\;\;i+j\leq k \label{rqr}
\ee
The solution of the system (\ref{rqr}) we shall use is 
\be
\alpha_i = i.
\ee

The dynamical realization of the above construction is in a phase space with 
coordinates
$\left\{q^0_a,\dots,q^{k-1}_b, p^0_a,\dots,p^{k-1}_b\right\}$ 
($a,b =1,\dots,c$).
The gauge transformation of the coordinate $q^i_a$ in this case is:
\be
\delta_\epsilon q^i_a = - \sum_{j=0}^i
{1\over j!}\epsilon^{(j)}_e q^{i-j}_b C_{eba} +
{1 \over i!}\epsilon_a^{(i+1)}
\ee
As it seen from the above equations,  the coordinates $q^0_a$ transform as
connection and $q^1$ transforms as $\dot q^0$.
Note that the gauge transformation of the Lagrange multipliers (\ref{varlm})
is of this type.

\section*{The Lagrangean with higher stage gauge symmetry}

Having a Hamiltonian $H$ and constraints $\psi_a$ it is possible to write down  
the following Lagrangean by analogy with the Lagrangean (\ref{lag}):
\be
L=\pp\dot \qq - H - \lambda_a \psi_a .\label{laghs}
\ee
In this Lagrangean the operators of the time derivatives
which are part of the definition of the constraints $\psi_a$
act on the Lagrangean multipliers $\lambda_a$.
It turns out that the Lagrangean (\ref{laghs}) is invariant under the
higher stage gauge transformation (\ref{gghs}) provided 
\be
\delta_\epsilon \lambda_a = \partial_t\epsilon_a  +
\epsilon_c C_{cba}\lambda_b. \label{mhs}
\ee
But this is exactly the gauge transformation of the Lagrange multipliers
in the standard $0$-stage case (see eq.(\ref{varlm})),
i.e. the gauge transformation of the Lagrange multipliers
does not depend on the stage of the gauge transformation.
Therefore,  the part of the constraints which acts in the Lagrange multipliers 
 phase space has an universal character and is given by eq.(\ref{lmc}).
This fact allows us to separate the contribution of the Lagrange multipliers
in the Lagrangean (\ref{laghs}).
Let $\hat q$ and $\hat p$ denote all phase space variables $\qq$ and $\pp$
but $\lambda$ and $\pi$. 
Then
the Lagrangean (\ref{laghs}) can be rewritten in the following form
\bea
L&=&\hat p\dot{\hat q} + \pi \dot\lambda
 - H - \lambda_a \hat\psi_a -\lambda_a(\dd \pi_a +
 \lambda_b C_{abc}\pi_c) \nn\\
 &=&\hat p\dot{\hat q} - H - \lambda_a \hat\psi_a\label{laghs2}
\eea
In eq.(\ref{laghs2}) $\hat\psi_a$ are the generators of the gauge symmetry
in the phase space $\{\hat q, \hat p\}$.
Note first, that there is no dependence on $\pi_a$ in $L$, so the
Lagrange multipliers are purely non-dynamical (as they should be). 
Second, eq.(\ref{laghs2}) describes a higher derivative model with gauge freedom.
The Hamiltonian approach to such models can be found in Ref.\cite{GT}.

\section*{Conclusion}

Eq. (\ref{cr}) shows that simple algebraic requirements
lead to very strong relations between 
the  constraints of different stage $\varphi^i_a$.
The structure of these constraints, as it is seen from eqs.(\ref{sky},\ref{lr}),
is dictated entirely from a representation $\pi$
of the constraint algebra $\ac$ and a number $k$.

Finally, we want to say few words about the applicability of our results
in the field models with gauge freedom.
In these models the gauge parameters
are functions not only of the time but of the spatial coordinates as well.
Therefore, the gauge transformation may depend on 
(higher) spatial derivatives of the gauge parameter.
In this case we can apply the procedure  
described above for the higher time derivatives to the spatial derivatives.
However, there is an essential difference between higher time derivatives and 
higher spatial derivatives in the Hamiltonian approach ---
the gauge transformations which depend on
the gauge parameter spatial derivatives are handled without any problem
in the Hamiltonian approach. 
Nevertheless, an analysis of the constraints  in the spirit of
eqs.(\ref{dec},\ref{cr}) and the reveal of the corresponding
algebraic structure (\ref{sky},\ref{lr}) seems instructive.

\end {document}